\documentclass{elektr}
\usepackage{hyperref}
\hypersetup{
colorlinks=true,
urlcolor=blue,
citecolor=blue}
\usepackage[all]{xy,xypic}
\usepackage{amsfonts,amssymb,amsmath,amsgen,amsopn,amsbsy,theorem,graphicx,epsfig}
\usepackage{eufrak,amscd,bezier,latexsym,mathrsfs,eurosym,enumerate}
\usepackage[utf8]{inputenc}
\usepackage[english]{babel}
\usepackage{cleveref,multirow}
\usepackage[dvipsnames]{xcolor}
\usepackage[pagewise]{lineno}
\usepackage{float}
\usepackage{graphicx}
\usepackage{filecontents}

\yil{}
\vol{}
\fpage{}
\lpage{}
\doi{}

\title{An Efficient End-to-End Deep Neural Network for Interstitial Lung Disease Recognition and Classification}

\author[M. S. Junayed, A. A. Jeny, M. B. Islam, I. Ahmed, and AFM S. Shah]{
\textbf{Masum Shah Junayed$^{1,2}$, Afsana Ahsan Jeny$^{1}$, Md Baharul Islam$^{1,3}$\thanks{bislam.eng@gmail.com}~, Ikhtiar Ahmed$^{4}$, A. F. M. Shahen Shah$^{5}$}\\
$^{1}$Department of Computer Engineering, Bahcesehir University, Istanbul, Turkey \\
$^{2}$Department of Computer Science \& Engineering, Daffodil International University, Dhaka 1207, Bangladesh \\
$^{3}$College of Data Science \& Engineering, American University of Malta, Bormla 1013, Malta\\
$^{4}$Department of Data Science, TU Dortmund University, 44227 Dortmund, Germany \\
$^{5}$Department of Electronics and Communication Engineering, Yildiz Technical University, Istanbul, Turkey\\ [1.8em]

\rec{.201}
\acc{.201}
\finv{..201}
}

\def\E{\ifmmode{\mathbb E}\else{$\mathbb E$}\fi} 
\def\N{\ifmmode{\mathbb N}\else{$\mathbb N$}\fi} 
\def\R{\ifmmode{\mathbb R}\else{$\mathbb R$}\fi} 
\def\Q{\ifmmode{\mathbb Q}\else{$\mathbb Q$}\fi} 
\def\C{\ifmmode{\mathbb C}\else{$\mathbb C$}\fi} 
\def\H{\ifmmode{\mathbb H}\else{$\mathbb H$}\fi} 
\def\Z{\ifmmode{\mathbb Z}\else{$\mathbb Z$}\fi} 
\def\P{\ifmmode{\mathbb P}\else{$\mathbb P$}\fi} 
\def\T{\ifmmode{\mathbb T}\else{$\mathbb T$}\fi} 
\def\SS{\ifmmode{\mathbb S}\else{$\mathbb S$}\fi} 
\def\DD{\ifmmode{\mathbb D}\else{$\mathbb D$}\fi} 

\newcommand{\bse}{\begin{subequations}}
\newcommand{\ese}{\end{subequations}}
\newcommand{\ben}{\begin{enumerate}}
\newcommand{\een}{\end{enumerate}}
\newcommand{\bens}{\begin{enumerate*}}
\newcommand{\eens}{\end{enumerate*}}
\newcommand{\be}{\begin{equation}}
\newcommand{\ee}{\end{equation}}
\newcommand{\bea}{\begin{eqnarray}}
\newcommand{\eea}{\end{eqnarray}}
\newcommand{\baa}{\begin{eqnarray*}}
\newcommand{\eaa}{\end{eqnarray*}}
\newcommand{\bc}{\begin{center}}
\newcommand{\ec}{\end{center}}

\theoremstyle{plain}

\theoremstyle{plain}

\theoremstyle{plain}

\theoremstyle{plain}

\theoremstyle{plain}

\theoremstyle{plain}

\theoremstyle{plain}

\theoremstyle{plain}

\begin{document}

\maketitle

\begin{abstract}
The automated Interstitial Lung Diseases (ILDs) classification technique is essential for assisting clinicians during the diagnosis process. Detecting and classifying ILDs patterns is a challenging problem. This paper introduces an end-to-end deep convolution neural network (CNN) for classifying ILDs patterns. The proposed model comprises four convolutional layers with different kernel sizes and Rectified Linear Unit (ReLU) activation function, followed by batch normalization and max-pooling with a size equal to the final feature map size well as four dense layers. We used the ADAM optimizer to minimize categorical cross-entropy. A dataset consisting of 21328 image patches of 128 CT scans with five classes is taken to train and assess the proposed model. A comparison study showed that the presented model outperformed pre-trained CNNs and five-fold cross-validation on the same dataset. For ILDs pattern classification, the proposed approach achieved the accuracy scores of 99.09\% and the average F score of 97.9\% that outperforms three pre-trained CNNs. These outcomes show that the proposed model is relatively state-of-the-art in precision, recall, f score, and accuracy.

\end{abstract}

\section{Introduction}\label{1}

The Interstitial Lung Diseases (ILDs) refer to a vast collection of respiratory disorders that include more than 200 conditions that affect the interstitial lung tissues \cite{troy2020diagnostic}. The inflammation and scar refer to pulmonary fibrosis that occurs because of lung tissue damage. ILD disrupts the normal process of oxygen transmission into the blood storm. Interrogating the patient about their physical examination, an actual intense assessment, aspiratory work testing, a chest X-beam, and a Computed Tomography (CT) filter are used to diagnose an ILD. The most appropriate convention for discriminating intra-class variation among specific patterns, as well as various combinations of diseased patterns, is high-resolution computed tomography (HRCT)\cite{depeursinge2012building}.

ILD is a histologically heterogeneous gathering of sicknesses, comparative clinical indications with one another, or even with various lung abnormalities, so differential determination is genuinely troublesome in any event for experienced doctors. Even for some accomplished specialists in the field, this characteristic property of ILDs can be complex, and there can be as much as 50\% inconsistency in radiological evaluation \cite{sluimer2006computer}. It is likewise tedious and arduous work and requires a specialist to anatomize an enormous number of cases. On the other hand, conducting a medical diagnostic introduces the patient to many dangers and raises healthcare expenditures, and such approaches may not always result in a solid diagnosis. To avoid these problematic issues, a statistical tool that integrates digital image processing and pattern recognition techniques, merged into computer-aided diagnosis (CAD) systems, has been widely researched to aid radiologists \cite{van2010comparing}. The most crucial parts of a CAD system are effective classification and recognition algorithms for various tissues. As a result, a wide range of traditional image descriptors and classifiers have been employed in the research.

\textbf{Problem Statement and Motivation.} Some researchers already worked for detecting and classifying lung diseases \cite{gupta2019evolutionary, anthimopoulos2014classification, sen2020depth, anthimopoulos2016lung, hattikatti2017texture, huang2020deep, bermejo2020classification}. Machine learning-based algorithms such as k-nearest neighbors (KNN) \cite{korfiatis2009texture}, Bayesian \cite{xu2006computer}, random forest (RF) \cite{anthimopoulos2014classification}, support vector machine (SVM) \cite{li2013lung}, and so on were used in their works. However, these works achieved less accuracy and computationally expensive due to inappropriate feature selection. Recently, Deep Learning (DL) techniques have already shown superior performance in various computer vision issues, raising hopes in the area, including medical image processing \cite{mossa2021ensemble}, classification \cite{junayed2020eczemanet}, segmentation \cite{hooda2019lung, uslu2020peri}, and detection \cite{junayed2019acnenet, jeny2020sknet}. Moreover, specific DL approaches called the pre-trained model of convolutional neural network (CNN), such as AlexNet \cite{krizhevsky2012imagenet}, VGG-16 \cite{simonyan2014very}, ResNet-50 \cite{he2016deep}, were used to achieve competitive accuracy. During training and testing, these works are time-consuming due to having deeper architectures with many parameters. Computerized Tomography (CT) scan images are used to classify ILD patterns through CNN-based architectures \cite{bermejo2020classification, anthimopoulos2016lung}. Small classes, more parameters, and fewer data are the key obstacles for having lower accuracy when using less efficient architectures for these papers. In \cite{huang2020deep}, ILD pattern classification images are used to classify through deep CNN-based architecture, but the suggested technique for ILD pattern categorization has certain drawbacks, which is much less effective than the slice-wise or conceptual segmentation approaches. He et al. \cite{he2020lung} also tried to classify the ILD patterns using the same customized CNN-based network, but their accuracy was too less because the features description is inadequate owing to the limited amount of feature maps, and overfitting has occurred in the result.

\textbf{Novelty and Contributions.}  Motivated by the above observations, in this article, we propose an architecture of deep convolution neural networks (DCNNs) for the classification of ILDs patterns using a relatively large and balanced image dataset including five classes: healthy (H), ground-glass opacity (GG), emphysema (EM), micronodules (MN), and fibrosis (FB). As an alternative to individually specifying a set of features \cite{depeursinge2011lung} and \cite{depeursinge2012near}, we developed a completely automated neural-based machine learning system to generate discriminant features from training data while also performing categorization. Because our approach is not problem-specific, it may be implemented in any other imaging area with relative ease. As opposed to an unsupervised RBM neural network \cite{li2013lung}, CNN is trained with the help of a supervised learning algorithm, and as a consequence, higher classification methods are anticipated. Furthermore, since lung images do not have unique visual features and only a limited amount of training examples, we modified the basic CNN design to address these problems. Adopting a single convolutional layer design and incorporating random neural node drop-out helped us decrease the number of parameters in the proposed CNN model. Data augmentation techniques such as translation, flip, rotation, scaling, shading, cropping, and affine transformations are applied to address the training dataset problem, which helped overcome the overfitting issue. To make our proposed model more robust and evaluate the performance, the pre-trained models such as AlexNet \cite{krizhevsky2012imagenet}, VGG-16 \cite{simonyan2014very}, and ResNet-50 \cite{he2016deep} are developed and compared. This paper's key contribution is listed below:

\begin{enumerate}
    \item An end-to-end deep Neural Network designed for ILD pattern classification. The network architecture is lightweight by utilizing fewer layers without compromising the performance of ILD pattern recognition and classification.
    
    \item We used seven different types of augmentation methods to increase the dataset size and non-overlapping ILDs patches extracted from each ROI \cite{gao2018holistic}.
    
    \item To demonstrate the high capability of the proposed method, three large pre-trained  CNNs,  i.e.VGG-16 \cite{simonyan2014very}, ResNet-50 \cite{he2016deep}, and AlexNet \cite{krizhevsky2017imagenet} have been tested on same dataset. 
\end{enumerate}

The following is how the rest of the paper is organized: Section \ref{2} discussed the most related works in literature: conventional approaches and deep learning approaches, Section \ref{3} depicts the proposed model for automatically classifying ILDs patterns, Section \ref{4} presents the material with implementation details, Section \ref{5} presents the findings and discussions of the model performance, and finally, we conclude our contributions in Section \ref{6}.

\section{Related Works}
\label{2}
There were several techniques to utilize CT images for computer-aided analysis and automated identification and classification of ILD. In this section, we briefly discuss them. 

\subsection{Traditional Approaches} 
For ILD Pattern characterization, a machine-based algorithm for automatic classification of a patient’s lung disease has been proposed. In \cite{gupta2019evolutionary}, they have focused on two major lung diseases, chronic obstructive pulmonary disease (COPD) and fibrosis. The authors applied several machine learning-based classifiers (kNN, SVM, RF, and DT for the feature selection method. Besides, three different algorithms are used to get better accuracy and to reduce the cost size. The proposed algorithm provides the novel accuracy where the improvised grey wolf algorithm (IGWA) belongs to the highest accuracy, but they used only two classes of lung disease. Based on local spectral analysis and random forest classification, Anthimopoulos et al. \cite{anthimopoulos2014classification} suggested a model for identifying lung tissue with disorders. With an average F-score of 89\%, their experimental results demonstrate improved performance and efficiency. 

Geraldo et al. \cite{ramalho2014lung} implemented a novel method for feature extraction of ACACM segmented images. They compare the discrimination capability of the proposed lung image descriptions to the usage of ELMNN. The SIM descriptors analyze the structural information where it is discriminate the pulmonary disease based on CT images. The proposed implemented method accomplished the 96\% accuracy to demonstrate the effectiveness of lung disease as COPD and fibrosis. However, this method needs to more robust automatic segmentation. Sen et al. \cite{sen2020depth} investigated to foresee the lung diseases with K-fold cross-validation and specially used five machine learning algorithms including Bagging, LR, RF, Logistic Model Tree, and Bayesian Networks. The accuracy of these algorithms was 88\%, 88.92\%, 90.15\%, 89.23\%, and 83.69\%, respectively. It shows that the RF algorithm gives the most elevated accuracy. Obtaining real-time data and comparing their outcomes with other data set was the main obstacle in their work.

\subsection{Deep Learning-based Approaches}
A few works have been done to recognize ILDs design order in large-scale color photos utilizing a deep CNN. In crafted by \cite{anthimopoulos2016lung}, they carried out deep CNN to group lung CT pictures for seven classes. 
To assess the model, they utilized 14696 pictures. 
The CAD framework for ILDs proposed a traditional element that incorporates three phases: (a) lung division, (b) lung illness measurement, and (c) differential conclusion with the 2D surface. A comparative investigation of the model demonstrated that it gives the expected accuracy of 85.5\% to dissect the lung design. The proposed drew nearer can be effectively-prepared on various textural lung patterns. Due to having countless boundaries to prepare the model, the entire preparing measure was generally sluggish, influencing the eventual outcome of the model. Pratiksha et al. \cite{hattikatti2017texture} proposed a texture-based lung disease classification on the 2D image. The deep learning classifier produced better accuracy than the machine learning classifier (e.g., SVM) 94\% and 87\%, respectively. The drawback of the paper is to use a fewer number of CT images was not sufficient to get the best discriminative results.

Agarwala et al. \cite{agarwala2020deep} utilized a fully CNN network to detect the ILD patterns analyzing a chest HRCT segment. To acquire database-specific characteristics, the pre-trained model was built using the well-known PASCAL VOC database and fine-tuned using the MedGIFT ILD database, which was used to fine-tune the model. They used only three lung disease pattern images, and Consolidation performed less well in terms of classification accuracy than either fibrosis or emphysema, but it did do better in terms of sensitivity and success rate than either of the other two diseases. In the MedGIFT ILD database, the success rate was 85.33\%, and in the PASCAL VOC database, it was only 78.67\% on average. It has happened as a result of a scarcity of lung disease pattern images.

In \cite{he2020lung}, a deep convolutional neural network is used to recognize lung nodules, and this technique is presented here (DCNN). Based on the features and difficulty of lung computed tomography (CT) pictures, the DCNN is developed. Based on the DCNN that has been built, the impacts of various model parameters, model structure, and optimization methods on classification results are examined. They used the same dataset as ours, and their picture size is only $32\times32$ pixels per slice in their experiment. However, owing to the limited number of feature maps and insufficiency of the feature description, the models' accuracy was substantially decreased, and the final model was overfitted.

In \cite{huang2020deep}, they developed a novel deep CNN model specifically for ILD pattern categorization. They also presented a new two-stage transfer learning (TSTL) technique to overcome the training data issue, which applies information acquired from adequate textural source data and supplementary unidentifiable lung CT data to the given dataset. An unsupervised approach with a goal function of forecast confidence was used to acquire unlabeled data. However, their proposed deep CNN technique for ILD pattern categorization is far less effective than the slice-wise or semantic segmentation methods.

Bermejo et al. \cite{bermejo2020classification} proposed a technique to distinguish and group radiographic examples of ILD at the essential stage in CT pictures. They principally utilized a multi-model troupe of deep CNNs involved 7 clusters, joining 2D, 2.5D, and 3D organizations. The outfit results accomplished a better (around 94\% accuracy) than every individual model, showing the probability of different classifier combinations. The examination recommended that their strategy be carried out on patients to recognize radiographic examples of ILD at a beginning phase but didn't deal with very similar pictures of ILD patterns.

\begin{figure*}[ht]
    \centering
    \includegraphics[width=0.95 \textwidth]{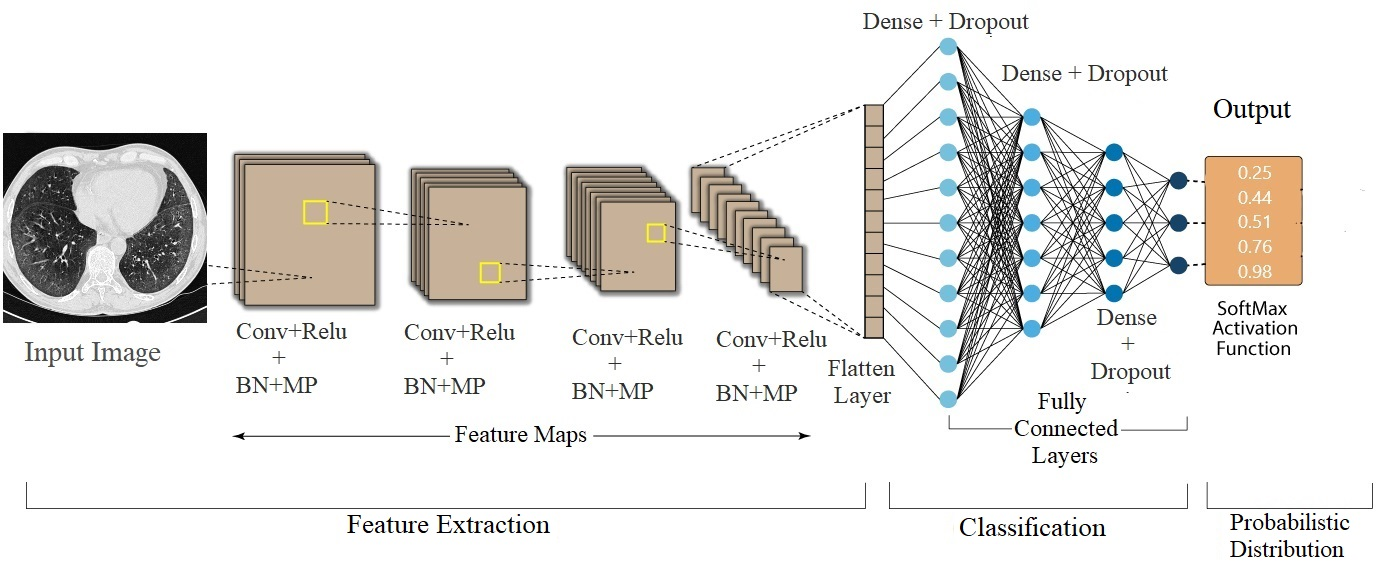}
    \caption{The proposed deep CNN architecture for ILD pattern identification and classification. An input image goes through 4 convolutional (Conv), batch normalization (BN), max-pooling (MP) layers, trailed by smoothing with three fully connected layers. The softmax classifier gives the last predictions to the ILDs five classes, including healthy (H), ground-glass opacity (GG), emphysema (EM), micronodules (MN), and fibrosis (FB).}
    \label{fig:architecture}
\end{figure*}

\section{Proposed Architecture}\label{3}

The CNNs have made essential contributions to image recognition and comprehension. It has gained popularity as a technique for deciphering medical images \cite{junayed2019acnenet, ramnarine2021line, che2021multi, jeny2020sknet} as a result of the emergence of new network variants and the introduction of reliable parallel solvers tailored for modern GPUs. It has been used successfully in various medical image processing, analysis, and classification applications by researchers \cite{dutta2021densely, khan2020classification,  junayed2020eczemanet, martinez2019comparison, shuvo2020lightweight}. Local textural elements containing complex, high-level structures with a particular orientation are used to categorize ILD patterns in CT images instead of random artifacts. We designed the deeper and more sophisticated CNN that can be trained using advanced computational tools, minimize computation time, and reducing parameter size.

\begin{table}[ht]
\centering
\caption{The layers, formation of layers, output form, and parameters of the recommended model are described in depth.}
\scalebox{0.9}{
\begin{tabular}{|c|c|c|c|}
\hline
\textbf{Layer} & \textbf{Formation of Layers} & \textbf{Output Form} & \textbf{Parameter} \\
\hline
\hline
     Input layer & $32\times32\times3$ & $32\times32\times3$ & - \\
\hline
     Conv2D & $32\times32\times3$; KS: $7\times7$; filters: 32; activation : ReLU & (None, 32, 32, 32) & 4736 \\
\hline
     BN & - & (None, 32, 32, 32) & 128 \\
\hline
     MP (2D)  & $2\times2$ & (None, 16, 16, 32) & 0 \\
\hline
     Conv2D & $16\times16\times32$; KS: $5\times5$; filters: 64; activation : ReLU & (None, 16, 16, 64) & 51264 \\
\hline
     BN  & - & (None, 16, 16, 64) & 256 \\
\hline
     MP (2D)  & $2\times2$ & (None, 8, 8, 64) & 0 \\
\hline
     Conv2D & $8\times8\times64$; KS: $3\times3$; filters: 32; activation : ReLU & (None, 8, 8, 96) & 55392 \\
\hline
     BN  & - & (None, 8, 8, 96) & 384 \\
\hline
     MP (2D)  & $2\times2$ & (None, 4, 4, 96) & 0 \\
\hline
     Conv2D & $4\times4\time96$; KS: $3\times3$; filters: 128; activation : ReLU & (None, 4, 4, 128) & 110720 \\
\hline
     BN  & - & (None, 4, 4, 128) & 512 \\
\hline
     MP (2D)  & $2\times2$ & (None, 2, 2, 128) & 0 \\
\hline
     Flatten  & N/A & (None, 512) & 0 \\
\hline
     Dense  & unit : 1024; activation : ReLU & (None, 1024) & 525312\\
\hline
     Dropout  & 0.25 & (None, 1024) & 0 \\
\hline
     Dense  & unit : 512; activation : ReLU & (None, 512) & 524800 \\
\hline
     Dropout  & 0.40 & (None, 512) & 0 \\
\hline
     Dense  & unit : 256; activation : ReLU & (None, 256) & 131328 \\
\hline
     Dropout  & 0.40 & (None, 256) & 0 \\
\hline
     Dense  & unit : 5; activation : 'Softmax' & (None, 5) & 1285 \\
\hline
\end{tabular}}
\label{table:details}
\end{table}

The proposed architecture is illustrated in Figure \ref{fig:architecture}. It has two stages for ILD pattern classification: feature learning and classification. The first stage includes convolutional, batch-normalization (BN), and max-pooling (MP) layers, while the fully connected layers and softMax activation function are used in the second stage. The input layer ($32\times32$ image size) is the same as the extracted image patch. It uses to preprocess data before feeding it into a neural network. It refers to converting an image to a vector and normalizing it to speed up training. The first convolution layer generated a $7\times7$ kernel size (KS) and 32 filters. It designs to create a feature map through the convolution of a series of weighted filters. The convolution operation is a point-multiplication summation of two-pixel matrices, (a) input data matrix, and (b) filter/feature matrix. Since different filters are used to obtain different feature maps, the other three convolution layers use 64, 32, and 128 filters. The KS for the second layer is $5\times5$, and the last two layers are set to $3\times3$.  Each of the convolution layers is attached with the ReLU. It is a non-linear function often used to speed up training because it always returns the maximum value. By adding some noise to the activation of the considered layer, BN is used to remove internal covariate shifts and induce regularization effects. Four MP layers with stride two are employed as a subsampling operation to filter the features in the sensing domain and extract the most important properties in the area. As a result, the output feature scale can be reduced while maintaining translation invariance and the number of parameters required.

The outputs of all of those layers are flattened and connected by three pairs of dense and dropout layers called fully connected layers. Every neuron in a fully connected layer is linked to every neuron in the next layer, resulting in output based on the entire image. The three dense layers of 512, 256, and 128 neurons are followed by 25\%, 40\%, and 40\% dropout. Finally, The softmax activation function measures the classification probability of lung disease patterns. It also normalizes feature maps in the range of (0, 1). Table \ref{table:details} demonstrates the configuration of the proposed model. The following is the softmax activation function:

\begin{equation} \label{eq:0}
    S(Z)_i = \frac{e^{z_i}}{\sum_{j=1}^K e^{z_j}}
\end{equation}
where, $S(Z)_i$ is the softmax input vector, $K$, $e^{z_i}$, and $e^{z_j}$ are the number of classes of ILD, standard input vector, and output vector, respectively.

The cross-entropy loss function is used in our proposed architectures. It is the most widely used loss function in CNN architectures. In ILDs pattern recognition and classification, categorial cross-entropy has been used to perform much better in eliminating outliers. This loss function is calculated to compute the following sum:

\begin{equation} \label{eq:1}
    Loss = - \sum^{output size}_{i=1} z_i . log y_i
\end{equation}
where, $z_i$ is the corresponding target value, and the output size is the number of scalar values, $y_i$ is the $i$th scalar value in the model output.

\begin{figure*}[ht]
    \centering
    \includegraphics[width=0.65\textwidth]{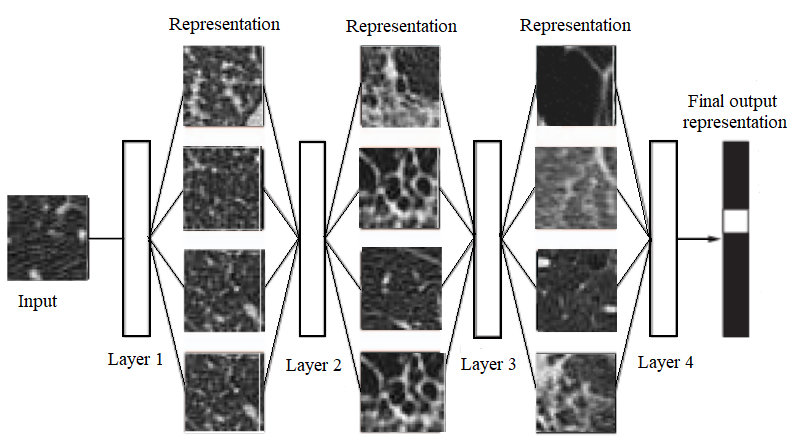}
    \caption{Feature representation of every convolution layers}
    \label{fig:feature}
\end{figure*}

Figure \ref{fig:feature} shows the ILD patch representations with many patch characteristics and learns more complex representations in subsequent levels. All convolutional layer representations are shown in this figure. For feature extraction, we utilized four convolution blocks. To deal with complicated data structures, deep learning often breaks down complex structures into simpler ones. As pictures have a unique spatial characteristic, the usage of layers is especially beneficial for patches. The visual qualities include sharp edges, contours, lines, textures, gradients, orientation, and color.

\section{Materials and Implementation Details}\label{4}

\subsection{Dataset}
\begin{figure*}[ht]
    \centering
    \includegraphics[width=0.65\textwidth]{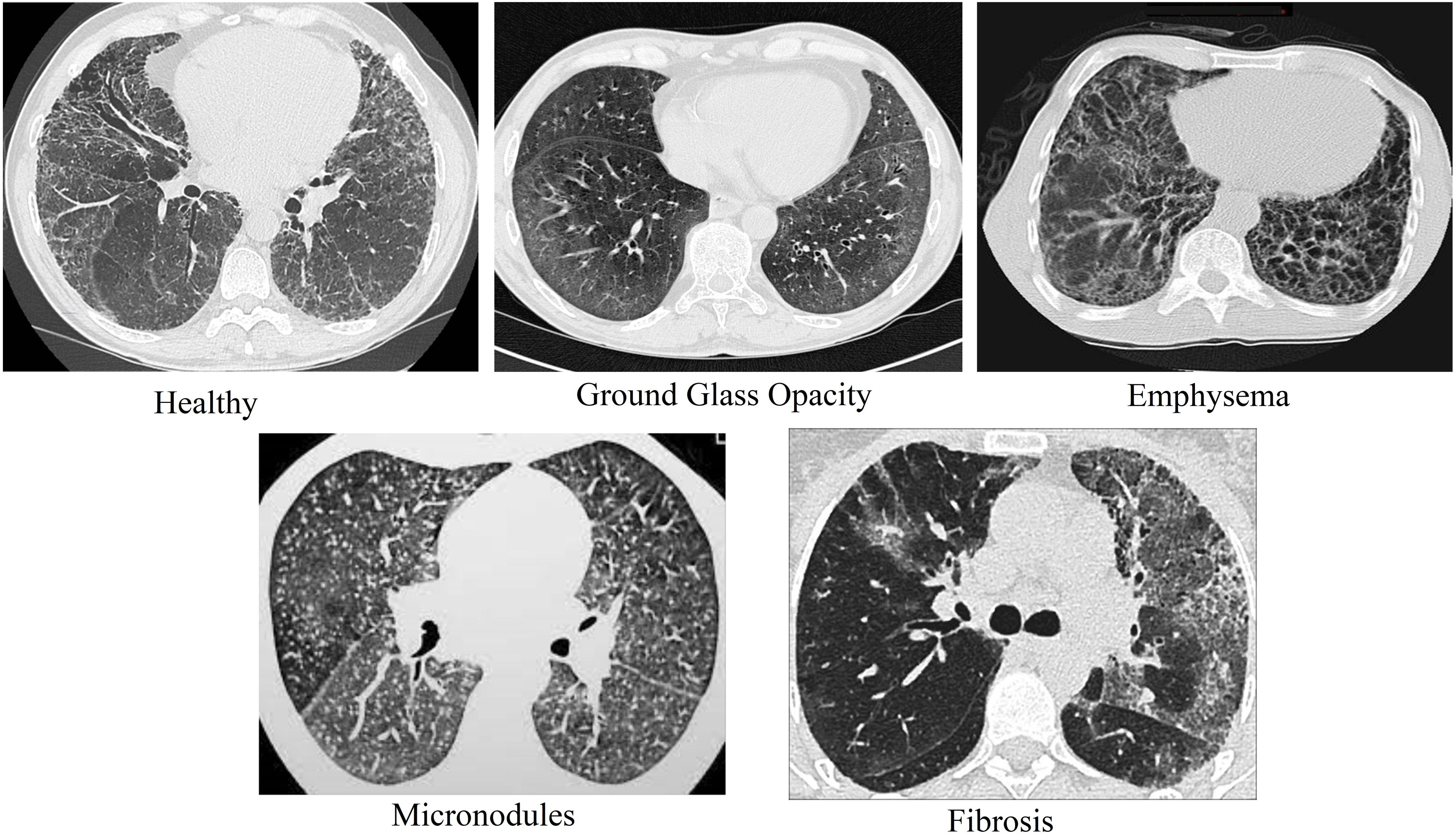}
    \caption{Sample of the dataset with five ILD classes.}
    \label{fig:data}
\end{figure*}

In this study, we used the publicly accessible interactive HUG database \cite{depeursinge2012building}, which contains data from 128 patients who underwent the HRCT scans at the University Hospitals of Geneva. It is composed of 109 annotated HRCT image series of $512\times512$ pixels per slice used to treat various ILD cases. This database is often manually annotated by professional radiologists, who use polygons to demarcate pathological patterns. There are seventeen distinct lung patterns and 1946 ROIs and clinical criteria from patients with histologically confirmed ILD diagnoses. The top five ILD patterns are chosen from all of the patterns. The considered classes are healthy (H), ground-glass opacity (GG), emphysema (EM), micronodules (MN), and fibrosis (FB). These are the most common patterns in this dataset, so we used these five ILDs groups in this study. Figure \ref{fig:data} portrays an example of the dataset.

\subsection{Preprocessing}

To standardize examines with various pixel separating, all pivotal cuts are rescaled. For the input of the proposed model, 3 input channels are generated using three different Hounsfield Unit (HU) windows. Following these serial preprocessing steps \cite{gao2018holistic}, based on the annotated polygons of these cases, a total of 9760 non-overlapping image patches with a size of 32 are extracted from each ROI. For picture patches, at any rate, 80\% of pixels falling inside the clarified polygonal districts, and 1383 patches are selected for each class. For image patches, at least 80\% of pixels falling inside the annotated polygonal regions, and we picked 783 patches for each class. Each class received 150 patches at random for testing, with 750 (approximately 20\%) patches in the test set. Also, the quantity of patches chosen depends on the fixed number of the most extraordinary class, which leaves approximately 80\% of the patches (3046) available for training and validation. To better generalize and avoid overfitting, the proposed model required more data and as much variation in the dataset as possible. As a result, seven transformations with label preservation, such as translation, flip, rotation, scaling, shading, cropping, and affine transformations, have been used to increase training data. Finally, we received 21328 training patches and used those patches for training and validation in this research.

\subsection{Implementation Setup}

Experiments run on a computer with the following specifications: Intel Core i9-10850K CPU with 3.60 GHz, 32 GB of RAM, and  NVIDIA Geforce RTX 2070 GPU with 8GB. Python, Keras, and Tensorflow environments were used to implement the suggested architecture and other previously trained CNN models. The pre-trained models are fine-tuned on the given dataset using augmentation methods while keeping the pre-trained weights and all levels of the model frozen. The pre-trained models are trained and validated using ImageNet weights. Three CNN models, such as VGG16 \cite{simonyan2014very}, ResNet-50\cite{he2016deep}, and AlexNet \cite{krizhevsky2017imagenet}, are already pre-trained on the ImageNet \cite{deng2009imagenet} dataset. The fully connected layers were removed from all of them, and a dense layer with 1024 neurons with ReLU activation functions and an output layer with five neurons with the softmax activation function was added in their stead. However, the suggested model has been trained for 32 batch size and reducing cross-entropy using Adam \cite{kingma2014adam} optimization to speed up learning by doing so for 50 epochs with the learning rate set to 0.00001.

\subsection{Hyperparameter Tuning}

To see how the amount of convolution blocks and optimizers (ADAM and MSE) impacts the accuracy of the classification and, as a result, the ILD recognition, three different models based fine-tuning on 3, 4, and 5 blocks with three different sets of filters, i.e. (16, 32, 64),(32, 64, 32, 128), and (32, 64, 32, 64, 128), respectively, are developed and evaluated on the dataset. Table \ref{table:convo} represents the accuracy achieved from these models. It is proved that the model's efficiency is improved while increasing the number of blocks at 4. On the other hand,  this efficiency is again decreased in the model at 5 blocks. Therefore, it is observed that the model with 4 blocks and (32, 32, 64, 128) filters outperforms the others.

\begin{table}[htb]
\centering
\caption{Accuracy analysis over the number of the blocks and different optimizers.}
\scalebox{0.9}{
\begin{tabular}{|c|c|c|c|c|} 
\hline
    Num. of the conv blocks & Description of filters & Optimizers & Accuracy (\%) & $F_s$ (\%) \\
\hline
    \multirow{2}{*}{3} & \multirow{2}{*}{16,32,64} & ADAM & 97.95 & 95.83 \\ 
     & & MSE & 97.35 & 95.26 \\
\hline
    \multirow{2}{*}{4} & \multirow{2}{*}{32,64,32,128} & ADAM & \textbf{99.09} & \textbf{97.98} \\
     & & MSE & 98.22 & 96.48 \\
\hline
    \multirow{2}{*}{5} & \multirow{2}{*}{32,64,32,64,128} & ADAM & 97.59 & 95.60 \\
     & & MSE & 96.99 & 95.07 \\
\hline
\end{tabular}}
\label{table:convo}
\end{table}

\subsection{Evaluation Criteria}
To classify ILD patches, a 5-fold cross-validation scheme is used in the evaluation process. The database is divided into five subsets randomly, with each one being used for training and testing.  The confusion matrix (CM), average F-score ($F_s = \frac{2* (Recall * Precision)}{(Recall + Precision)}$), accuracy ($\frac{TP + TN}{TP + TN + FP + FN}$), recall (sensitivity) ($\frac{TP}{TP + FN}$), and precision ($\frac{TP}{TP + FP}$) are utilized to assess the performance across the various classes. The CM is a metric for evaluating the efficiency of a classification model that is frequently represented as a matrix. It includes four measurements: true positive (TP), true negative (TN), false positive (FP), and false-negative (FN) that quantify machine learning efficiency, as well as the problems of deep learning classification. The achievement of the suggested architecture is assessed the $E_{avg}$ utilizing the following equations (\ref{eq:4}).

\begin{equation}\label{eq:4}
    F_{avg} = \frac{1}{N}\sum ^{N}_{c=1} F_s
\end{equation}

where $N$ denotes the number of classes while $F_s$ denotes the F Score for $c$ classes.

\begin{figure}[ht]
    \centering
    \includegraphics[width=0.65\textwidth]{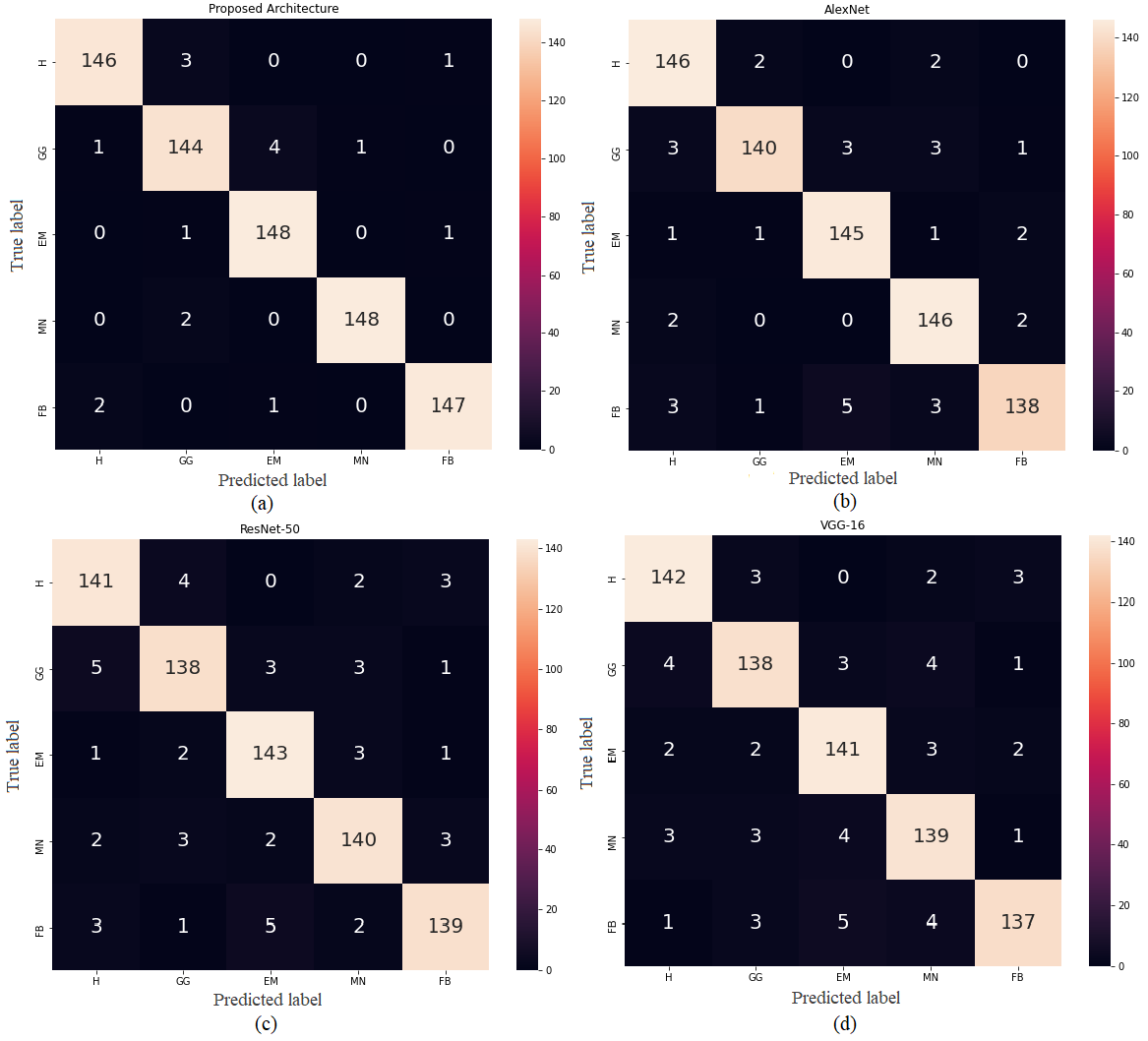}
    \caption{The CM of the suggested model and three pretrained CNNs. Here, (a) Proposed model, (b) AlexNet, (c) ResNet-50, and (d) VGG-16.}
    \label{fig:cm}
\end{figure}

\section{Experimental Results}\label{5}
\subsection{Performance}
Figure \ref{fig:cm} depicts the CM of 5 classes of ILD patterns in CT images, which is a visual representation of each classifier's (proposed model, AlexNet, ResNet-50, and VGG-16) efficiency. The diagonal values address the correct arrangement number for each class, while the others address the number of disorders between each of the two classes. The CM of the proposed model yielded more true positive values than the CMs of the other pre-trained models (AlexNet, ResNet-50, and VGG-16). Furthermore, we can see that each class of the CM is predicted subtly, with the values of each class being 146, 144, 148, 148, and 147, respectively.
\begin{table}[t]
\centering
\caption{This table depicts the performance assessment of the suggested model and various pre-trained models. In terms of Accuracy, Precision, Recall, and F1-score, the VGG-16, ResNet-50, AlexNet, and the suggested technique were compared to each of the ILD pattern classes.}
\scalebox{0.8}{
\begin{tabular}{|c|c|c|c|c|c|c|}
\hline
\textbf{Method} & \textbf{ILD Classes} & \textbf{No. of Truth} & \textbf{Accuracy (\%)} & \textbf{Recall (\%)} & \textbf{Precision (\%)} & \textbf{F1 -Score($F_{s})$ (\%)} \\ \hline
\multirow{5}{*}{VGG-16} & Healthy & 152 & 97.60 & 93.42 & 94.67 & 94.04 \\
 & Ground Glass & 149 & 97.06 & 92.62 & 92.62 & 92.63 \\
 & Emphysema & 153 & 97.20 & 92.16 & 94.00 & 93.06 \\
 & Micronodules & 152 & 96.80 & 91.45 & 92.67 & 92.05 \\
 & Fibrosis & 144 & 97.33 & 95.14 & 91.33 & 93.20 \\ \hline
\multicolumn{3}{|c|}{Total Average} & 97.19 & 92.96 & 93.06 & 92.99 ($F_{avg}$) \\ \hline
\multirow{5}{*}{ResNet-50} & Healthy & 152 & 97.33 & 92.76 & 94.00 & 93.38 \\
 & Ground Glass & 148 & 97.07 & 93.24 & 92.00 & 92.62 \\
 & Emphysema & 153 & 97.73 & 93.46 & 95.33 & 94.39 \\
 & Micronodules & 150 & 97.33 & 93.33 & 93.33 & 93.33 \\
 & Fibrosis & 147 & 97.47 & 95.56 & 92.67 & 93.60 \\ \hline
\multicolumn{3}{|c|}{Total Average} & 97.39 & 93.67 & 93.47 & 93.46 ($F_{avg}$) \\ \hline
\multirow{5}{*}{AlexNet} & Healthy & 155 & 98.47 & 94.19 & 97.33 & 98.74 \\
 & Ground Glass & 144 & 98.13 & 97.22 & 93.33 & 95.24 \\
 & Emphysema & 153 & 98.27 & 94.27 & 96.67 & 95.71 \\
 & Micronodules & 155 & 98.27 & 94.19 & 97.33 & 95.74 \\
 & Fibrosis & 143 & 97.33 & 96.50 & 92.00 & 94.20 \\ \hline
\multicolumn{3}{|c|}{Total Average} & 98.09 & 94.47 & 95.33 & 95.93 ($F_{avg}$) \\ \hline
\multirow{5}{*}{Proposed Model} & Healthy & 149 & 99.07 & 97.33 & 97.95 & 97.66 \\
 & Ground Glass & 150 & 98.40 & 96.01 & 96.03 & 96.00 \\
 & Emphysema & 153 & 99.07 & 96.73 & 98.67 & 97.69 \\
 & Micronodules & 149 & 99.60 & 99.33 & 98.67 & 99.00 \\
 & Fibrosis & 149 & 99.33 & 98.66 & 98.00 & 98.33 \\ \hline
\multicolumn{3}{|c|}{Total Average} & \textbf{99.09} & \textbf{97.61} & \textbf{97.86} & \textbf{97.98 ($F_{avg}$)} \\ \hline
\end{tabular}}
\label{table:performn}
\end{table}

Table \ref{table:performn} demonstrates the performance comparison of the suggested model with the pre-trained models where it can be observed that the number of truth values of the pre-trained models aren't always close except the proposed model. For example, VGG-16 is classified the truth values are 152, 149, 153, 152, and 144 respectively, from 5 classes and the values are not always familiar. However, the provided values in the proposed model are 149, 150, 153, 149, and 149, indicating that the classification rate is relatively high and similar for each class. In the case of accuracy, all the classes of VGG-16, ResNet-50, and AlexNet are provided the accuracy between 97\% to 98\% almost (VGG-16: 97.19\%, ResNet-50: 97.39\%, and AlexNet: 98.09\%). But the model improves the accuracy at (99.09\%). The proposed model also achieved the optimal classification performances (Recall: 97.61\%, Precision: 97.86\%, F1-Score: 97.98\%) than VGG-16 (Recall: 92.96\%, Precision: 93.06\%, F1-Score: 92.99\%), ResNet-50 (Recall: 93.67\%, Precision: 93.47\%, F1-Score: 93.46\%), and AlexNet (Recall: 94.47\%, Precision: 95.33\%, F1-Score: 95.93\%) respectively.

\begin{figure}[ht]
    \centering
    \includegraphics[width=0.85\textwidth]{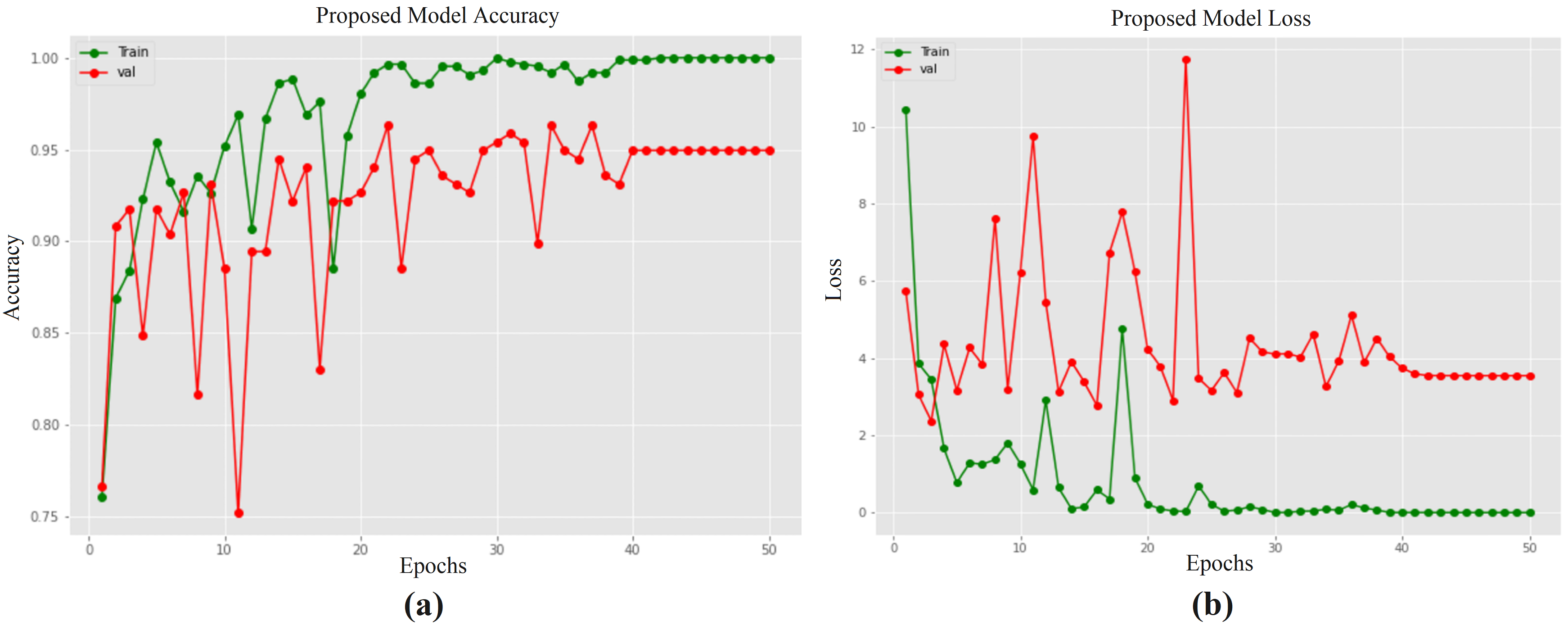}
    \caption{The proposed architecture's accuracy (a) and loss (b) graph. Training is shown by the green line, while validation is shown by the red line. Each epoch is represented by the X-axis, while the Y-axis represents the accuracy, and the loss respectively.}
    \label{fig:loss}
\end{figure}

Figure \ref{fig:loss} (a) and (b) show the accuracy (train vs validate) as well as the loss (train vs validate) graph of the proposed model. With regards to graph figure \ref{fig:loss}  (a), both TA and VA are showed some fluctuations until hitting 40 epochs then both peaked at 99.86\% and 95.17\% in 50 epochs respectively. On the other hand, in the case of figure \ref{fig:loss} (b), some oscillations are occurred in the TL and VL before falling to 40 epochs then both gathered 0.08\% and 5.48\% loss in 50 epochs respectively. According to the figure, the training loss decreases when the epoch number increases, while training accuracy and validation accuracy increase. Validation loss decreases from 6.72\% to 3.66\% between the period 10 and 40 epoch. It is slightly increased (5.48\%) at 40 to 50 epochs. Following the explanation of the 50 epochs from figure \ref{fig:loss}, the final loss of training, loss of validation, accuracy of training, and accuracy of validation are 0.08\%, 99.86\%, and 95.17\% respectively. Therefore, the proposed system performed exceptionally well, with higher accuracy and negligible losses that is able to show an competitive performance.

\subsection{K-fold Cross Validation}

K-fold cross-validation is used on the entire dataset to validate the model further to investigate if 5-fold cross-validation performs better than the 80-20\% split for training and testing data. It is accomplished by partitioning the whole dataset into five subsets, four of which were used to train the model and one for testing. This procedure is continued until all the subsets are utilized once as testing.


\begin{figure}[ht]
    \centering
    \includegraphics[width=0.75 \textwidth]{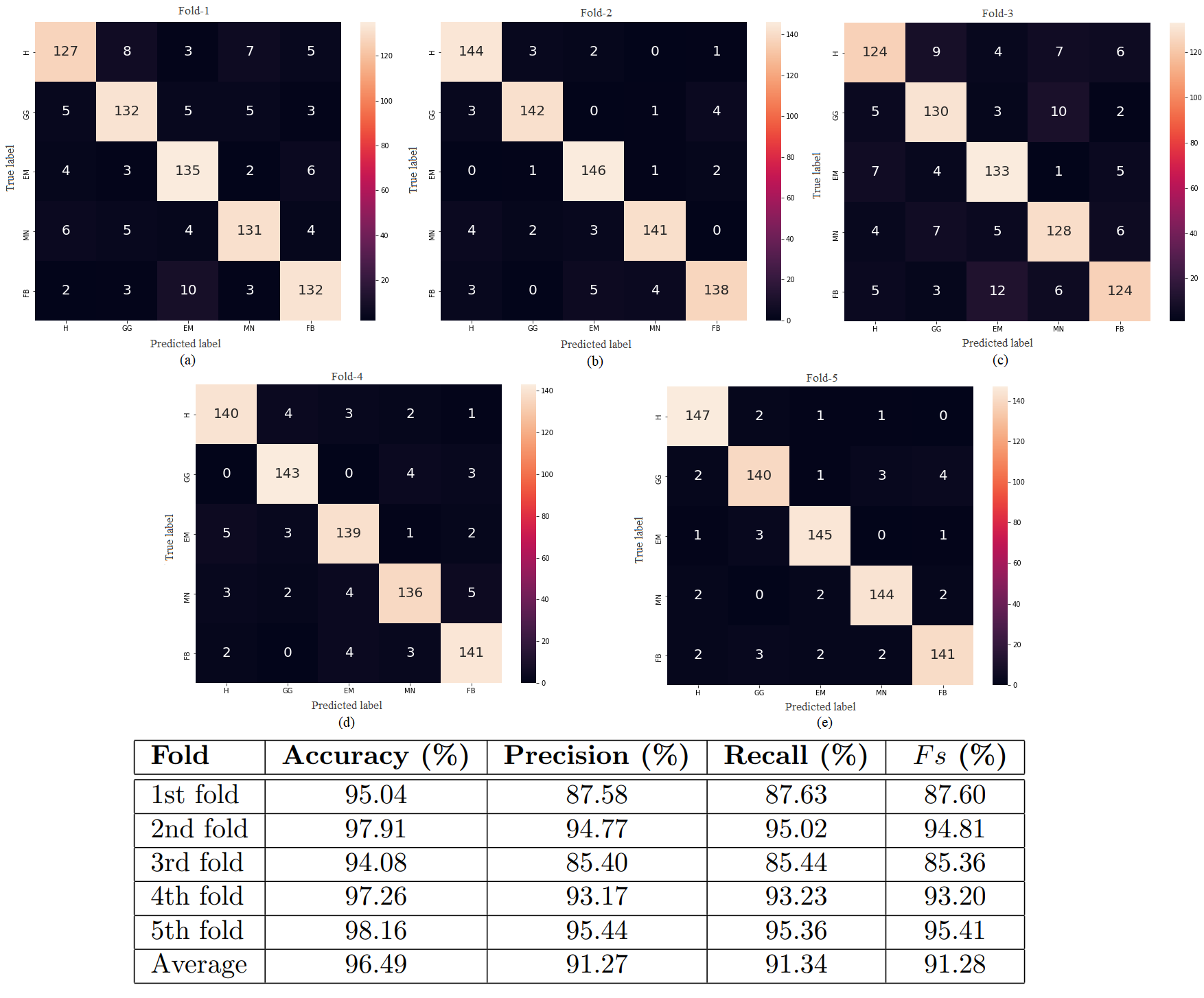}
    \caption{The performance of the Confusion Matrix of cross-validation process for each fold: (a) fold 1, (b) fold 2, (c) fold 3, (d) fold 4, and (e) fold 5 where H = Healthy, GG = Ground Glass, EM = Emphysema, MN = Micronodules, and FB = Fibrosis. In this figure, the performance measures in terms of confusion matrix, accuracy, precision, recall, and f-1 score.}
    \label{fig:foldcm}
\end{figure}
Test results of the proposed model for ILD patterns in the CT image dataset extend the 5-fold cross-validation confusion metrics are presented in Figure \ref{fig:foldcm}. All the performances of the fold are quite good. Fold 5 has a far higher classification rate than the other folds, with all of its values exceeding 140. Similarly, except for the Fibrosis class, all other classes can provide 140 classification rates in fold 2. In 4-fold, the values of the two classes are less than 140 and the rest are up. On the other hand, all classes of fold 1 are capable of displaying more than 130 values except the Healthy class. However, the rate of incorrect classification of fold 4 is significantly higher than that of other folds. This figure also displays the performance of each fold. The average precision, accuracy, recall, and F1-Score of 5 folds are 96.49\%, 91.27\%, 91.34\%, and 91.28\% respectively. Therefore, we conclude that the proposed model does not display any variation in each fold, ensuring reliability and universality. However, the 10-fold cross-validation is also applied on the dataset. After the experiment, we observed that 5-fold outperformed the 10-fold cross-validation. Due to the lack of space, we displayed only the result of 5-fold cross-validation.

\subsection{Comparison to the state of the arts}

\begin{table}[ht]
\centering
\caption{Statistics for comparison of different existing deep learning (DL) based approaches with our proposed work in terms of DL methods, $F_{avg}$, accuracy, sensitivity and precision The values in bold correspond to the best performance.}
\scalebox{0.9}{
\begin{tabular}{|c|c|c|c|c|c|c|} 
\hline
  \textbf{Approaches} &  \textbf{DL Methods} & \textbf{$F_{avg}$ (\%)} & \textbf{Accuracy (\%)} & \textbf{Sensitivity (\%)} & \textbf{Precision (\%)} \\
 \hline
   Li et al. \cite{li2014medical} (2014) & CNN & - & - & 77.40 & 74.00 \\
\hline    
    Anthimopoulos et al. \cite{anthimopoulos2016lung} (2016) & Deep CNN & 85.47 & 85.61 & - & - \\
\hline
    Wang et al. \cite{wang2017multiscale} (2017) & MRCNN & 89.56 & 90.10 & 89.64 & 89.50 \\
\hline
    He et al. \cite{he2020lung} (2020) &  Deep CNN & 66.70 & 66.80 & - & - \\
\hline
    Huang et al. \cite{huang2020deep} (2020) &  Deep CNN & 96.74 & - & - & - \\
\hline
    \textbf{Proposed Work} & Deep CNN & \textbf{97.98} & \textbf{99.09} & \textbf{97.61} & \textbf{97.86} \\
\hline
\end{tabular}}
\label{table:pwork}
\end{table}

Table \ref{table:pwork} compares a fair comparison among our suggested model and the state-of-the-art approaches that employed the same dataset and a variety of customized deep CNN classifiers. In the presented comparison, it can be observed that Li et al. \cite{li2014medical} obtained 77.4\% sensitivity and 74\% precision using CNN, which is 20.21\% and 23.86\% lower than ours, respectively, simply because their CNN was not able to distinguish different visual structure when it was applied to ILD images. On the other hand, he et al. \cite{he2020lung} achieved the lowest accuracy (66.8\%) and f1-score (66.7\%) as a result of the deep CNN structure used in their model. Due to their structure's inability to appropriately divide the training and testing sets, the limited amount of feature maps, they are experiencing an overfitting issue in their data. As of MRCNN \cite{wang2017multiscale}, the batch size was too small, the training was insufficient, and the parameter adjustments were inadequate, resulting in a lower recognition rate overall. Anthimopoulos et al. \cite{anthimopoulos2016lung} used less convolutional kernels in their deep CNN structure. Their accuracy is 13.48\% less than ours (99.09 vs. 85.61). Huang et al. \cite{huang2020deep} did not provide their accuracy, and the f1-score is only calculated at 1.24\% less than ours. They claimed that due to their database itself, they got fewer results. Finally, we also utilized a deep CNN structure and obtained a high outcome than others in this ILD pattern classification because we have attempted to combine the size of the training set and choose the suitable batch size in order to guarantee that each parameter modification is based on sufficient training and backpropagation data for each parameter.

\section{Conclusions and Future Works}\label{6}

This article introduces an end-to-end deep CNN model for classifying Interstitial Lung Disease (ILD) patterns using lung CT image patches. After the preprocessing step, the publicly available HUG database is used with five ILD patterns, including healthy tissues. A new architecture is developed to classify captures of low-level textural characteristics of lung tissue. This architecture consists of four convolutional layers with various filters, ReLU activation, batch normalization, max-pooling, and flatten, dense, and dropout layers. Categorical cross-entropy is employed as a loss function for training purposes, and adam is also used to optimize the model. The proposed approach achieved an accuracy of 99.09\% that is higher than the AlexNet, ResNet-50, and VGG-16 by absolute margins of 1.00\%, 1,81\%, and 1.61\%, respectively. Furthermore, the suggested system outperformed or is comparable to current state-of-the-art precision, F1-score, recall, and accuracy. It would be interesting to develop an automatic system that classifies ILDs patterns and classifies three-dimensional CT scans and integrates different ILD diagnosis techniques as a supportive tool for radiologists with very few computational parameters in the future.

\section*{Experimental Code Availability}
For further investigation, comparison, analysis of this study by the research community, the experimental code and model are accessible upon request to the corresponding author through email.

\bibliographystyle{vancouver}
\bibliography{refs}
\end{document}